\documentclass[twocolumn, aps, pra, preprintnumbers, amsmath, amssymb, showpacs]{revtex4}%

\usepackage{graphicx}
\usepackage{dcolumn}
\usepackage{bm}

\begin{document}

\title{Two-photon wave mechanics}

\author{Brian J. Smith}
 \email{bsmith4@uoregon.edu}

\author{M. G. Raymer}

\affiliation{Oregon Center for Optics and Department of Physics, University of Oregon, Eugene, 
Oregon 97403}

\date{\today}

\begin{abstract}
The position-representation wave function for multi-photon states and its equation of motion are introduced. A major strength of the theory is that it describes the complete evolution (including polarization and entanglement) of multi-photon states propagating through inhomogeneous media. As a demonstration of the two-photon wave function's use, we show how two photons in an orbital-angular-momentum entangled state decohere upon propagation through a turbulent atmosphere.
\end{abstract}

                              
\maketitle

There are two approaches to solving problems in quantum mechanics: quantum field theory (QFT), and wave mechanics (WM). In WM, the fundamental physical entities are ``particles'', whose collective state is described by a wave function. To treat few-particle systems in WM, such as the helium atom with two electrons, one formulates and solves the two-electron wave equation in position space. In QFT, the fundamental physical entities are fields, which are decomposable into modes, each of which can have various numbers of excitations. 
It is clear that there is a one-to-one correspondence between \it modes \rm in QFT, and \it states \rm in WM.

Each approach has its realm of preferred applicability. For example, one does not usually treat the helium atom with QFT, but one does use this theory when treating high-energy electron collision experiments. 
Just as one most often uses WM in atomic physics, we advocate the use of this approach for few-photon phenomena, such as those encountered in elementary quantum information schemes (quantum cryptography, one-way quantum computing, and linear optics quantum computation). However, as far as we know, a complete photon-wave-mechanics (PWM) theory of electromagnetism has not been introduced. Development of the PWM description of multiple photons leads toward completion of the quantum description of electromagnetism, which must include entanglement and decoherence not yet treated.

In this paper, we briefly review the one-photon wave function in coordinate space, of which there are several proposed \cite{landau_peierls, birula_94, birula_96, sipe, hawton_95, hawton_99, dragoman_06}. Choosing one of these, we introduce a two-photon wave function, along with its tensor equation of motion, which we call the two-photon Maxwell-Dirac equation. Its relationships to other well-known formulations, such as quantum electrodynamics (QED) of few-photon wave packets \cite{titgla}, and vector-field classical coherence theory \cite{mandel_wolf} are then brought to light. 
These connections suggest that we have chosen the most appropriate single-photon formalism upon which to base a generalization to multiple photons. Choosing a different single-photon formalism does not give these close relations and leads to different equations of motion, normalization conditions, and Hilbert-space scalar products. We also demonstrate the use of the two-photon wave function in a calculation of quantum-state disentanglement for a pair of spatially-entangled photons traveling through the atmosphere. The two-photon formalism is readily extended to multiple-photon states. The theory gives a lucid view of the photon as a particle-like quantum object, making a pedagogical link between standard quantum wave mechanics and quantum field theory, which must be equivalent according to current understanding.

Much of the confusion surrounding the definition of the photon wave function in the position representation arises from the non-localizability of the photon and the corresponding absence of a position operator \cite{newton_wigner, birula_98}. This is due to the fact that the photon has zero mass, and is a spin-1 object, with only two independent spin degrees of freedom. Of the coordinate-space single-photon wave functions proposed \cite{landau_peierls, birula_94, birula_96, sipe, hawton_95, hawton_99, dragoman_06}, we find the most useful is the Birula-Sipe formulation, defined in terms of the localization of photon energy \cite{birula_94,birula_96,sipe}  
\begin{equation}
\Psi ^{\left( 1 \right)} \left( {{\bf{x}},t} \right) = \left[ {
\begin{array}{*{20}c}
   {\phi _{ + 1}^{\left( 1 \right)} \left( {{\bf{x}},t} \right)}  \\
   {\phi _{ - 1}^{\left( 1 \right)} \left( {{\bf{x}},t} \right)}  \\
\end{array}} \right],
\label{eq:1}
\end{equation}
rather than, for example, the Landau-Peierls non-local number density wave function \cite{landau_peierls}. Here $\phi_{\sigma}^{\left( 1 \right)}$ is a three component, complex vector labeled by $\sigma = \pm 1$ for positive(negative) helicity \cite{birula_94,birula_96}. In vacuum this wave function satisfies the Dirac-like equation \cite{birula_94, birula_96, sipe, raymer_smith_05, scully_05}
\begin{equation}
i\hbar \partial _t \Psi ^{\left( 1 \right)} =
\hat H\,\Psi ^{\left( 1 \right)}
= \hbar c\Sigma _3 \nabla  \times \Psi ^{\left( 1 \right)},
\label{eq:2}
\end{equation}
and the zero-divergence condition $\nabla\cdot\Psi^{\left(1\right)}=0$. Here $\Sigma_3$ is a 
Pauli-like matrix that changes the sign of the negative helicity component in Eq.(1).
The curl and divergence operators are understood to act on the upper and lower components of $\Psi^{\left(1\right)}$ separately, $\hbar$ is Planck's constant (which cancels in Eq.(\ref{eq:2})), and $c$ is the speed of light in vacuum. The single-photon Hamiltonian is $\hbar c\Sigma_3\nabla\times$. Equation (\ref{eq:2}) and the zero-divergence condition are formally equivalent to the classical Maxwell equations, as can be seen by substituting
\begin{equation}
\phi_\sigma^{\left(1\right)}\left({{\bf{x}},t}\right)
= \frac{{{\bf D}^{\left(+\right)} \left( {{\bf x},t} \right)}}{{\sqrt {2\epsilon_0 } }} + i\sigma \frac{{{\bf B}^{\left(  +  \right)} \left( {{\bf x},t} \right)}}{{\sqrt {2\mu_0 } }},
\label{eq:3}
\end{equation}
where $\bf{D}^{\left( + \right)}$ and $\bf{B}^{\left( + \right)}$ are the positive-frequency parts of the electric-displacement and magnetic-induction fields, and $\epsilon_0$ ($\mu_0$) is the vacuum permittivity (permeablity).

In a linear, isotropic medium, $\epsilon_0$ and $\mu_0$ in Eq.(\ref{eq:3}) are replaced by spatially dependent functions $\epsilon ({\bf{x}})$ and $\mu ({\bf{x}})$, and the equation of motion Eq.(\ref{eq:2}), is changed by the material interaction to \cite{birula_94}
\begin{equation}
i\hbar \partial _t \Psi ^{\left( 1 \right)} =
\hat H\,\Psi ^{\left( 1 \right)}
= \hbar v\Sigma _3 \left( \nabla + \nabla L \right) \times \Psi ^{\left( 1 \right)},
\label{eq:med}
\end{equation}
where the speed of light is constructed from the local values of permittivity and permeability in the medium, $ v ({\bf{x}}) = 1/\sqrt {\epsilon({\bf{x}}) \mu({\bf{x}}) } $, and the matrix $L$ has the form,
\begin{equation}
L({\bf{x}}) = \left({\openone \ln \sqrt {\epsilon({\bf{x}}) \mu({\bf{x}}) }  + \Sigma _1 \ln \sqrt {\epsilon({\bf{x}}) /\mu({\bf{x}}) } } \right)/2.
\label{eq:L}
\end{equation}
Here $\openone$ is the identity matrix and $\Sigma_1$ is a Pauli-like matrix that interchanges the two helicity components of Eq.(\ref{eq:1}). In a medium the photon Hamiltonian is given by $\hbar\, v\Sigma_3\left(\nabla+\nabla L\right)\times$.
The modified divergence condition in a medium is $\left(\nabla+\nabla L\right)\cdot \Psi^{\left(1\right)}=0$.

The integrated square modulus of the photon wave function over all space gives the expectation value of the photon's energy
\begin{equation}
\int{\Psi^{\left( 1 \right)} \left({{\bf{x}},t}\right)^\dag \Psi^{\left( 1 \right)}\left({{\bf{x}},t}\right)d^3 x}
= \left\langle {E_1 } \right\rangle.
\label{eq:4}
\end{equation}
One can associate with this wave function a local probability density 
$\rho\left({{\bf{x}},t}\right)=\Psi^{\left(1\right)}\left({{\bf{x}},t}\right)^\dag
\Psi ^{\left(1\right)}\left({{\bf{x}},t}\right)/\left\langle{E_1 }\right\rangle $,
and current density
$j\left({{\bf{x}},t}\right)
=\Psi^{\left(1\right)}\left({{\bf{x}},t}\right)^\dag{\bf{s}}
\Psi^{\left(1\right)}\left({{\bf{x}},t}\right)/\left\langle {E_1}\right\rangle$,
that obey the continuity equation \cite{birula_94,birula_96}. Here $\bf{s}$ is a vector
composed of the three spin-1 matrices. This probability density and current density are defined in relation to the photon energy, not photon number.

The appropriate scalar product is best formulated in momentum space, where a local photon-number probability density is well defined \cite{birula_96}. Transformed into the position representation, the scalar product is found to be a nonlocal integral, consistent with the absence of a local photon particle-density amplitude \cite{birula_96}. The fact that photon wave functions representing orthogonal states are not orthogonal with respect to an integral of the form Eq.(\ref{eq:4}) is consistent with the well-known nonexistence of localized, orthogonal spatio-temporal modes in QED \cite{titgla}.

We now propose that the two-photon wave function $\Psi^{\left(2\right)}\left({{\bf{x}}_1 ,{\bf{x}}_2 ,t}\right)$, which is related to the probability amplitude for finding the energies of two photons localized at two different spatial positions ${\bf{x}}_1$ and ${\bf{x}}_2$, at the same time $t$, with the photons in any polarization state, can be constructed from single-photon wave functions as
\begin{equation}
\Psi ^{\left(2\right)}\left({{\bf{x}}_1 ,{\bf{x}}_2 ,t}\right)
= \sum\limits_{l,m} {C_{lm} \,\psi _l^{\left( 1 \right)}\left({{\bf{x}}_1 ,t}\right)\otimes
\psi _m^{\left(1\right)}\left({{\bf{x}}_2 ,t}\right)},
\label{eq:5}
\end{equation}
where the coefficients $C_{lm}$, symmetrize the wave function, and $\otimes$ is the tensor product. The modulus squared of the coefficients $|C_{lm}|^2$, gives the probability of the photons being in the states labeled by $l$ and $m$. Each tensor component is related to the two-photon spin state's energy probability density. The basis states $\left\{{\psi _l^{\left( 1 \right)}}\right\}$, are solutions of the single-photon wave equations (Eq.(\ref{eq:2}) in free space and Eq.(\ref{eq:med}) in a linear medium), and include spin dependence. The equation of motion for the two-photon wave function is found by adding the Hamiltonians for the individual photons
\begin{eqnarray}
  i\hbar \partial _t \Psi ^{\left( 2 \right)} & = & \hbar v_1\alpha _1^{\left( 2 \right)} \left( {\nabla _1  + \nabla _1 L_1 } \right) \times \Psi ^{\left( 2 \right)} {} \nonumber\\
   & + & \hbar v_2\alpha _2^{\left( 2 \right)} \left( {\nabla _2  + \nabla _2 L_2 } \right) \times \Psi ^{\left( 2 \right)} ,
\label{eq:6}
\end{eqnarray}
where
$\alpha_1^{\left(2\right)} = \Sigma_3 \otimes \openone$, 
$\alpha_2^{\left(2\right)} = \openone \otimes \Sigma_3$,
the curl operators are understood to act on appropriate components of the tensor product, $L_{1(2)}=L({\bf{x}}_{1(2)})$, and $v_{1(2)}=v({\bf{x}}_{1(2)})$. In free space the $L$s drop out and the speed of light takes on its vacuum value $c$. We call equation (\ref{eq:6}) the Maxwell-Dirac equation for a two-photon state. The two-photon wave function also obeys the divergence conditions
\begin{equation}
\left( {\nabla _j  + \nabla _j L_j } \right) \cdot \Psi ^{\left(2\right)} = 0, \quad j = 1,2.
\label{eq:6div}
\end{equation}
Tracing over the tensor product of the two-photon wave function and its Hermitian conjugate,
and integrating over all space gives the expectation value of the product of the two photons' energies
\begin{equation}
\iint {Tr\left[ {\Psi ^{\left(2\right) \dag} \Psi^{\left(2\right)}} \right ]d^3 x_1 d^3 x_2 } = \left\langle {E_1 E_2 } \right\rangle.
\label{eq:7}
\end{equation}
If the state of the photons is not entangled, this equals
$\left\langle {E_1 E_2 } \right\rangle = 
\left\langle E_1 \right\rangle \left\langle E_2 \right\rangle$.
One can also define a joint probability density
$\rho^{\left(2\right)}\left({{\bf{x}}_1 ,{\bf{x}}_2 ,t}\right)
= Tr\left[\Psi^{\left(2\right)}\left({{\bf{x}}_1 ,{\bf{x}}_2 ,t}\right)^\dag
\Psi ^{\left(2\right)}\left({{\bf{x}}_1 ,{\bf{x}}_2 ,t}\right)\right ]/\left\langle{E_1 E_2 }
\right\rangle$,
for finding the energy of one photon at the space-time coordinate $ x_1$, and the other at $x_2$, ($ x_j = \left({{\bf{x}}_j,t}\right)$, $j=1,2$), and current density $j^{\left(2\right)}\left({{\bf{x}}_1 ,{\bf{x}}_2,t}\right)$, obeying a continuity equation.

To demonstrate the Birula-Sipe single-photon theory \cite{birula_94, birula_96, sipe} is best suited to PWM, we first show there is a direct relation between the $n$-photon wave function and the $n$-photon detection amplitude of quantum optics \cite{mandel_wolf, klyshko, scully_zubairy, pittman_96, walborn_2003}
\begin{equation}
 A _D^{\left( n \right)} \left( {{\bf{x}}_1 ,{\bf{x}}_2 , \cdots ,{\bf{x}}_n ;t} \right) = \left\langle vac \right|\mathop { \otimes \,}\limits_{j = 1}^n { \bf{\hat E}}^{(+)} \left( {{\bf{x}}_j ,t} \right)\left| {\Psi ^{\left( n \right)} } \right\rangle,
\label{eq:11}
\end{equation}
whose modulus-squared is proportional to the probability for joint, $n$-event detection. If one neglects the magnetic field, assuming the detectors respond only to the electric field, the effective $n$-photon wave function is just a tensor product of $n$ electric-field vectors evaluated at potentially different spatial values, exactly the same form as Eq.(\ref{eq:11}). Then the $n$-photon wave function can be identified with the spatial mode of the electromagnetic field, showing the connection between modes and states, and appealing to the choice of the single-photon formalism based on energy localization.

To further strengthen the case for the energy-density single-photon theory, we show the close connection of the two-photon wave function and classical coherence theory. Considering two positive-helicity photons for simplicity, the two-photon wave function can be written as a sum of four terms
\begin{eqnarray}
 \Psi ^{\left(2\right)}\left({x_1 ,x_2}\right)  =  \left[ {{\bf{D}}\left( {x_1 } \right) / \sqrt{2\epsilon ({\bf{x}}_1)} + i{\bf{B}}\left( {x_1 } \right) / \sqrt{2\mu ({\bf{x}}_1)}} \right] \nonumber\\
 \otimes \left[ {{\bf{D}}\left( {x_2 } \right) / \sqrt{2\epsilon ({\bf{x}}_2)} + i{\bf{B}}\left( {x_2 } \right) / \sqrt{2\mu ({\bf{x}}_2)}} \right] . \qquad
\label{eq:12}
\end{eqnarray}
Each term has the same form as one of the four second-order coherence matrices of classical coherence theory \cite{mandel_wolf} 
\begin{equation}
{\sf{A}}\left( {x_1 ,x_2 } \right) = \left\langle {{\bf{F}}^ *  \left( {x_1 } \right) \otimes {\bf{G}}\left( {x_2 } \right)} \right\rangle,
\label{eq:cma}
\end{equation}
which give a complete description of second-order partial coherence of an optical field (including spatial, temporal and polarization coherence). Here $ {\bf{F}},{\bf{G}} \in \{ {\bf{D}},{\bf{B}} \} $, and the brackets $\left\langle {} \right\rangle $ imply an ensemble average over all realizations of the fields. 
Evolution of these matrices is described by a set of linear differential equations \cite{mandel_wolf} that we call the first-order Wolf equations, which are equivalent to Eqs.(\ref{eq:6}) and (\ref{eq:6div}). This equivalence shows a deep connection between propagation of classical coherence quantities and multi-photon states. In addition, each component of the coherence matrices obeys the (second-order) Wolf equations \cite{mandel_wolf}, a well-known set of classical second-order differential equations recently highlighted for their relation to the two-photon detection amplitude \cite{saleh}. In much the same way that the Klein-Gordon equation does not completely describe the evolution of electron states by neglecting spin, the same holds for the second-order Wolf equations, which do not specify the relations between polarization components. In contrast, the new Maxwell-Dirac equation, Eq.(\ref{eq:6}), contains all such relationships. In this sense our result shows the quantum origin of the classical Wolf equations \cite{saleh}, and illuminates their connection to the propagation behavior of multi-photon states. Choice of a different single-photon theory upon which to base the multi-photon theory would not lead to the above results.

To illustrate the utility of the two-photon wave function and its relation to classical
coherence theory, we consider the propagation through a turbulent atmosphere of two
quasi-monochromatic photons, initially entangled in their spatial degrees of freedom, as depicted in Fig. 1. We assume the photons are emitted from a source in opposite directions occupying one of two orbital angular momentum (OAM) states, described by the Laguerre-Gauss wave functions $\psi_{p,l}\left({r,\theta}\right) = R_l^p\left(r\right)\exp\left({il\theta}\right)/\sqrt{2\pi}$, where $r$ and $\theta$ are cylindrical coordinates. Here $R_l^p \left(r\right)$ is the radial wave function, $\exp\left({il\theta}\right)/\sqrt{2\pi}$ is the angular wave function, $l$ is the OAM quantum number, and $p$ is the radial quantum number. 
\begin{figure}[b]
 \includegraphics[width = 2.14 in]{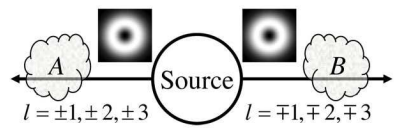}
 \caption{\label{Fig1} Diagram of a thought experiment involving two photons, initially entangled in their OAM states, $ l = \pm 1, \pm 2, \pm 3 $, traveling through independent, random phase atmospheres, labeled by $A$ and $B$.}
\end{figure}
We assume that both photons, labeled $A$ and $B$, have the same polarization, radial quantum number $p = 0$, and consider orbital quantum numbers of equal magnitudes $|l|$, separately. We take the two-photon basis as
\begin{equation}
\begin{array}{l}
 \varphi _{AB}^1  = \psi _{p,l}^A  \otimes \psi _{p,l}^B ,\quad \ {} \; {} \varphi _{AB}^2  = \psi _{p,l}^A  \otimes \psi _{p, - l}^B , \\ 
 \varphi _{AB}^3  = \psi _{p, - l}^A  \otimes \psi _{p,l}^B ,\quad \varphi _{AB}^4  = \psi _{p, - l}^A  \otimes \psi _{p, - l}^B , \\ 
 \end{array}
\label{eq:15}
\end{equation}
where $\psi _{p,l}^{A(B)}$ is evaluated at the coordinate of photon $A(B)$. For concreteness, we treat the input pure state $\Psi _{in}^{\left(2\right)} = [\varphi _{AB}^2 + \varphi _{AB}^3]/ \sqrt{2}$. The photons pass through independent, thin, dielectric, Gaussian phase-randomizing atmospheres, modeled by a quadratic phase structure function \cite{leader_78, shirai_2003}. This determines the form of the medium function $L({\bf{x}})$ in Eq.(\ref{eq:L}), which we solve in the paraxial approximation. The elements of the density matrix $\rho$, at the output of the turbulence are determined by integrating the radial power distributions $r\left|{R_l^{p = 0}\left(r\right)}\right|^2$, for each photon multiplied by the
circular-harmonic transform of the phase correlation function $C_\phi$, which describes the effect of the atmosphere on the state \cite{paterson}. In our model, the phase correlation function for each atmosphere is
\begin{eqnarray}
C_\phi ^{A\left( B \right)} \left({r_ ,\Delta\theta}\right) = 
\exp\left\{{ -\frac{1}{2}\left[{D_\phi ^{A\left(B\right)}\left({r,\Delta\theta}\right)}\right]}\right\},
\label{eq:16}
\end{eqnarray}
where
$D_\phi ^{A(B)}\left({r,\Delta\theta}\right)=
\left[{2r\sin\left({\Delta\theta}\right)/r_{A(B)}}\right]^2$
is the quadratic phase structure function \cite{shirai_2003} of the aberrations in atmosphere $A(B)$, and $r_{A\left(B\right)}$ is the transverse length scale of the corresponding turbulence. We find a closed form expression for the circular harmonic transform
\begin{equation}
\begin{array}{c}
 \tilde C_\phi ^{A\left( B \right)} \left( {r,m} \right) = \int_0^{2\pi } {C_\phi ^{A\left( B \right)} \left( {r,\Delta \theta } \right)\exp \left( { - im\Delta \theta } \right)d\Delta \theta }  \\ 
  = \exp \left[ { - 2\left( {r/r_{A(B)} } \right)^2 } \right]I_m \left[ {2\left( {r/r_{A\left( B \right)} } \right)^2 } \right] , \\ 
 \end{array}
\label{eq:17} 
\end{equation}
where $m$ is $0$ or $2l$, and $I_m\left(x\right)$ is an $m$th-order modified Bessel function.

Being interested only in two OAM states for each photon, $\psi_{p,\pm l}$, we
may treat each photon as a qubit. Other photon states can be considered to be loss channels \cite{mitchell_2003} and we normalize the post-selected density matrix. We examine the decay of entanglement by calculating the concurrence  $C(\rho)$ \cite{wootters}, from the normalized density matrix, as a function of the ratio of the optical beam waist to the characteristic turbulence length scale $w/r_0$ \cite{paterson}. Each atmosphere is assumed to have the same coherence length $r_0$. For a maximally entangled state, $C=1$, and for a non-entangled state, $C=0$. We assume the atmosphere is unmonitored, so any independent information about its fluctuations is lost, leading to loss of entanglement. We plot the concurrence in Fig. 2(a), for various initial OAM quantum numbers.

This result shows that for a beam waist much smaller than the turbulence length, $w << r_0$, the entanglement is more robust to the turbulent atmosphere. Physically this reflects the fact that the photons will experience few phase distortions across their wave fronts. These results also indicate that entangled states with larger OAM values experience less disentanglement through a turbulent atmosphere. This appears to be due to the fact that scattering from one OAM state to another depends only on the change in OAM, $\hbar \Delta l$ \cite{paterson}, which must be supplied by the atmosphere. The atmosphere can, on average, change the OAM of the light only by a particular amount set by the spatial fluctuations that characterize it. We also calculate the fidelity of the output two-photon state relative to the input state, which for a pure-state input is
\begin{equation}
F\left( {\rho _{in} ,\rho _{out} } \right) = \left\langle {\Psi _{in} } \right|\rho _{out} \left| {\Psi _{in} } \right\rangle.
\end{equation}
This result, plotted in Fig. 2(b) for the input state given above, indicates that states with small OAM values, and thus small ``rms'' beam width \cite{shirai_2003}, have higher overall transmission than do states with large OAM values. We should stress that the overall transmission of the OAM states depends on the beam waist $w$. We conclude that entangled states with smaller waists and larger OAM quantum numbers will be more robust to turbulence.
\begin{figure}[t]
 \includegraphics[width = 85.725 mm]{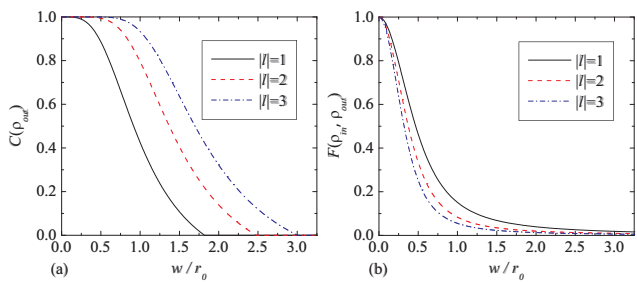}
 \caption{\label{Fig2} (Color online) (a) Concurrence as a function of the ratio, $w/r_0$, of beam waist to turbulence length scale for three different magnitudes of input OAM. (b) Fidelity of the output with respect to the input pure state.}
\end{figure}

We have introduced the two-photon wave function based on energy localization. This two-photon wave function obeys the two-photon Maxwell-Dirac equation, which is equivalent to the equations of motion of the classical second-order coherence matrices \cite{mandel_wolf}. The connections we have found between this wave function, QED wave-packet-mode detection amplitudes, and classical coherence theory give credence to the choice to use the energy-localization wave function rather than others, such as the Landau-Peierls wave function \cite{landau_peierls}, which is a non-local ``number density'' wave function. The formalism provides powerful tools to analyze the behavior of few-photon states, as shown by the example above, where we calculated the disentanglement of a spatially entangled two-photon state by using essentially classical field equations. This theory is well suited to the study of realistic implementations of linear optical quantum computing \cite{klm}, measurement-induced nonlinearities with linear optics \cite{sanaka_2006}, and continuous-variable entanglement through quantum state tomography \cite{smith_2005,lvovsky_raymer}. The well-defined Lorentz transformation properties of this wave function \cite{birula_96} make it ideal for the examination of relativistic quantum information with photons \cite{peres_terno}.

The authors wish to thank Iwo Bialynicki-Birula, Cody Leary, Jan Mostowski, John Sipe, and Davison Soper for helpful comments. This research was supported by the National Science Foundation, grant nos. 0219460 and 0334590.

\bibliography{prl_2006}
\end{document}